\documentclass{elsart}
\usepackage{graphics}
\begin{document}
\begin{frontmatter}
\title{Dyson--Schwinger calculations of meson form factors}
\author{Pieter Maris}
\address{Dept. of Physics, Kent State University, Kent  OH 44242}
\date{8 August 2000}
\begin{abstract}
The ladder-rainbow truncation of the set of Dyson--Schwinger equations
is used to study the pion and kaon electromagnetic form factors and the
\mbox{$\gamma^\star \pi^0 \gamma$} transition form factor in impulse
approximation.  With model parameters previously fixed by the
pseudoscalar meson masses and decay constants, the obtained form factors
are in good agreement with the data.
\end{abstract}
\end{frontmatter}
\section{The Dyson--Schwinger Equations}
%
Our aim is to obtain the hadron spectrum and observables such as
coupling constants and form factors from the underlying theory,
QCD\footnote{Supported by NSF grant No.~PHY97-22429 and computer
resources from NERSC.}.  The set of Dyson--Schwinger equations [DSEs]
form a useful tool for this purpose~\cite{dserev}.  In rainbow-ladder
truncation, they have been successfully applied to calculate the masses
and decay constants of light pseudoscalar and vector
mesons~\cite{MR97,MT99v}.  The dressed-quark propagator, as obtained
from its DSE, together with the Bethe--Salpeter amplitude and the
$qq\gamma$ vertex as obtained from the homogeneous and inhomogeneous
Bethe--Salpeter equations [BSE] respectively, form the necessary
elements for form factor calculations in impulse
approximation~\cite{MT00pi,MT00ka}.

The DSE for the renormalized quark propagator in Euclidean space is
\begin{equation}
\label{gendse}
 S(p)^{-1} = i \, Z_2\, /\!\!\!p + Z_4\,m(\mu)
        + Z_1 \int\!\!\frac{d^4q}{(2\pi)^4} \,g^2 D_{\mu\nu}(k) 
        \frac{\lambda^a}{2}\gamma_\mu S(q)\Gamma^a_\nu(q,p) \,,
\end{equation}
where $D_{\mu\nu}(k)$ is the dressed-gluon propagator,
$\Gamma^a_\nu(q;p)$ the dressed-quark-gluon vertex, and $k=p-q$.  The
most general solution of Eq.~(\ref{gendse}) has the form
\mbox{$S(p)^{-1} = i /\!\!\! p A(p^2) + B(p^2)$} and is renormalized 
at spacelike $\mu^2$ according to \mbox{$A(\mu^2)=1$} and
\mbox{$B(\mu^2)=m(\mu)$} with $m(\mu)$ the current quark mass.

The DSE for the $qq\gamma$ vertex \mbox{$\Gamma_\mu(p_+,p_-)$} is the
inhomogeneous BSE
\begin{equation}
 \Gamma_\mu(p_+,p_-) = Z_2 \, \gamma_\mu + 
        \int\!\!\frac{d^4q}{(2\pi)^4} \, K(p,q;Q) 
        \;S(q_+) \, \Gamma_\mu(q_+,q_-) \, S(q_-)\, ,
\label{verBSE}
\end{equation}
where $p_\pm = p\pm\frac{1}{2}Q$ are the incoming and outgoing quark
momenta, and similarly for $q_\pm$.  The kernel $K$ is the renormalized,
amputated $\bar q q$ scattering kernel that is irreducible with respect
to a pair of $\bar q q$ lines.  Solutions of the homogeneous version of
Eq.~(\ref{verBSE}) define vector meson bound states at timelike photon
momenta \mbox{$Q^2=-m_{\rm v}^2$}.  It follows that
$\Gamma_\mu(p_+,p_-)$ has poles at those locations.  Pseudoscalar
solutions $\Gamma_{\rm ps}(p_+,p_-;Q)$ of the homogeneous BSE define
bound states such as pions and kaons.  Together with the canonical
normalization condition for $q \bar q$ bound states, Eqs.~(\ref{gendse})
and (\ref{verBSE}) completely determine all elements needed for form
factor calculations in impulse approximation.

To solve the BSE, we use a ladder truncation, with an effective
quark-antiquark interaction that reduces to the perturbative running
coupling at large momenta~\cite{MT99v}.  In conjunction with the rainbow
truncation for the quark DSE, the ladder truncation preserves both the
vector Ward--Takahashi identity [WTI] for the quark-photon vertex and
the axial-vector WTI.  This ensures the existence of massless
pseudoscalar mesons connected with dynamical chiral symmetry
breaking~\cite{MR97}.  In combination with impulse approximation, it
also guarantees current conservation~\cite{MT00ka}.  Our model
preserves the one-loop renormalization group behavior of QCD and
reproduces perturbative results in the ultraviolet region.  As long as
the interaction is sufficiently strong in the infrared region, it leads
to chiral symmetry breaking and confinement.  The model gives a good
description of the $\pi$, $\rho$, $K$, $K^\star$ and $\phi$ masses and
electroweak decay constants~\cite{MT99v}, with only two parameters in
the effective interaction fitted to $f_\pi$ and the condensate, and
realistic current quark masses fitted to $m_\pi$ and $m_K$.
\section{Results for meson electromagnetic form factors}
\begin{figure}
\begin{center}
\resizebox{0.9\textwidth}{!}{
 \includegraphics{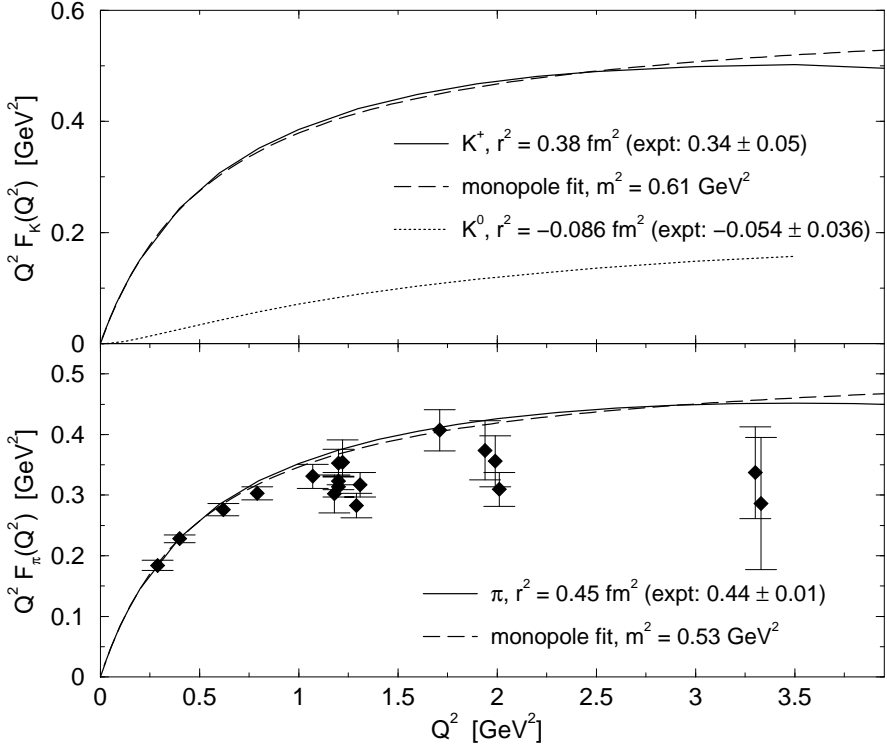}} \caption{\label{fig1} The electromagnetic
 form factors and radii of the light pseudoscalar mesons, adapted from 
 \cite{MT00ka}.  The pion data are from Ref.~\protect\cite{B78}, 
 the charge radii from \protect\cite{A86A86KM78}.
}
\end{center}
\end{figure}

%
Meson form factors in impulse approximation are described by two
diagrams, with the photon coupled to the quark and to the antiquark
respectively.  We can define a form factor for each of these diagrams,
e.g.
\begin{eqnarray}
 2\,P_\nu\,F_{a\bar{b}\bar{b}}(Q^2) &=&
        N_c\int\!\!\frac{d^4q}{(2\pi)^4}
        \,{\rm Tr}\big[ S^a(q) \, \Gamma_{\rm ps}^{a\bar{b}}(q,q_+;P_-) 
\nonumber \\ &&{}\times
        S^b(q_+) \, i \Gamma^{b}_\nu(q_+,q_-)\, S^b(q_-) \,
        \bar\Gamma_{\rm ps}^{a\bar{b}}(q_-,q;-P_+) \big] \;, 
\end{eqnarray}
where \mbox{$q = k+\frac{1}{2}P$}, 
\mbox{$q_\pm = k-\frac{1}{2}P \pm \frac{1}{2}Q$},  
\mbox{$P_\pm = P \pm \frac{1}{2}Q$}.  We work in the isospin symmetry
limit, so for the pion form factor we have \mbox{$F_{\pi}(Q^2) =
F_{u\bar{u}u}(Q^2)$}.  The charged and neutral kaon form factors are 
given by \mbox{$F_{K^+}= \frac{2}{3}F_{u\bar{s}u} + 
\frac{1}{3}F_{u\bar s \bar s}$} and \mbox{$F_{K^0} = 
-\frac{1}{3}F_{d\bar{s}d} + \frac{1}{3}F_{d\bar s\bar s}$} respectively.
Our results for $Q^2 F_\pi$, $Q^2 F_{K^+}$, and $Q^2 F_{K^0}$ are shown
in Fig.~\ref{fig1}, together with the corresponding charge radii.

\begin{figure}
\begin{center}
\resizebox{0.9\textwidth}{!}{
 \includegraphics{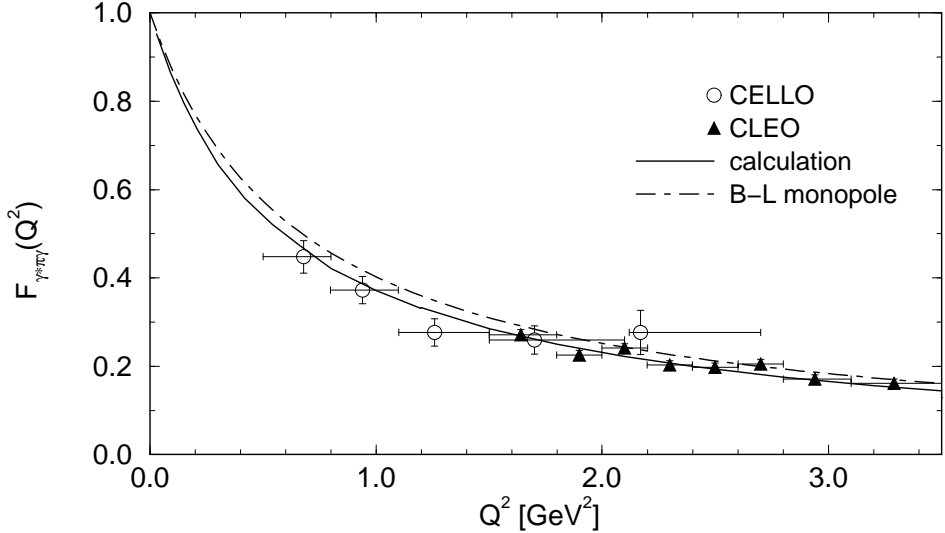}}
 \caption{\label{fig2} 
 The $\gamma^\star\,\pi\gamma$ form factor, with data from 
        CLEO and CELLO~\cite{cellocleo}.}
\end{center}
\end{figure}

The impulse approximation for the $\gamma^\star\pi\gamma$ vertex 
with $\gamma^\star$ momentum $Q$ is
\begin{eqnarray}
\lefteqn{ \Lambda_{\mu\nu}(P,Q)=i\frac{\alpha }{\pi f_{\pi }}
        \,\epsilon_{\mu \nu \alpha \beta }\,P_{\alpha }Q_{\beta }
        \, g_{\pi\gamma\gamma}\,F_{\gamma^\star\pi\gamma}(Q^2)  } \\
& & \nonumber
        =\frac{N_c}{3}\, \int\!\frac{d^4q}{(2\pi)^4}
        {\rm Tr}\left[S(q)\, i\Gamma_\nu (q,q')\,S(q')\, 
                i\Gamma_\mu (q',q'')\,S(q'')\,\Gamma_\pi(q'',q;P)\right] \;.
\end{eqnarray}
where the momenta follow from momentum conservation.  In the chiral
limit, the value at $Q^2 = 0$, corresponding to the decay \mbox{$\pi^0
\rightarrow \gamma\gamma$}, is given by the axial anomaly and its value
\mbox{$g^{0}_{\pi\gamma\gamma}=\frac{1}{2}$} is a direct consequence of 
only gauge invariance and chiral symmetry; this value is reproduced by
our calculations and corresponds well with the experimental width of
$7.7~{\rm eV}$.  In Fig.~\ref{fig2} we show our results with realistic
quark masses, normalized to the experimental $g_{\pi\gamma\gamma}$.

All our form factor results are remarkably close to the data, without
any readjustment of the parameters.  Up to about $Q^2 = 3\,{\rm GeV}^2$,
they can be fitted quite well by monopoles.  Asymptotically, our
calculated form factors behave like \mbox{$Q^2 F(Q^2) \rightarrow c$} up
to logarithmic corrections.  However, numerical limitations prevent us
from accurately determining these constants. Around $Q^2 = 3\,{\rm
GeV}^2$, our result for $Q^2 F_\pi$ is well above the pQCD result
\mbox{$16 \pi f_\pi^2 \alpha_s(Q^2)$}~\cite{FJ}, 
and clearly not yet asymptotic; $F_{\gamma^\star\pi\gamma}$ lies in
between monopoles fitted to the Brodsky--Lepage asymptotic limit, $8
\pi^2 f_\pi^2$~\cite{BL}, and to the DSE limit, 
$\frac{16}{3}\pi^2 f_\pi^2$~\cite{dsegpglimit}.
\end{document}